\documentclass[10pt,twocolumn,reprint,amsmath,amssymb,aps,pra,showpacs,longbibliography,superscriptaddress]{revtex4-1}
\usepackage[latin9]{inputenc}
\usepackage{amsmath}
\usepackage{amssymb}
\usepackage{graphicx}

\makeatletter
\usepackage{amsfonts}
\usepackage{graphicx}
\usepackage{graphics}

\makeatother

\begin{document}
\title{Heavy polarons in ultracold atomic Fermi superfluids at the BEC-BCS
crossover: formalism and applications}
\author{Jia Wang}
\affiliation{Centre for Quantum Technology Theory, Swinburne University of Technology,
Melbourne 3122, Australia}
\author{Xia-Ji Liu}
\affiliation{Centre for Quantum Technology Theory, Swinburne University of Technology,
Melbourne 3122, Australia}
\author{Hui Hu}
\affiliation{Centre for Quantum Technology Theory, Swinburne University of Technology,
Melbourne 3122, Australia}
\date{\today}
\begin{abstract}
We investigate the system of a heavy impurity embedded in a paired
two-component Fermi gas at the crossover from a Bose-Einstein condensate
(BEC) to a Bardeen--Cooper--Schrieffer (BCS) superfluid via an extension
of the functional determinant approach (FDA). FDA is an exact numerical
approach applied to study manifestations of Anderson\textquoteright s
orthogonality catastrophe (OC) in the system of a static impurity
immersed in an ideal Fermi gas. Here, we extend the FDA to a strongly
correlated superfluid background described by a BCS mean-field wavefunction.
In contrast to the ideal Fermi gas case, the pairing gap in the BCS
superfluid prevents the OC and leads to genuine polaron signals in
the spectrum. Thus, our exactly solvable model can provide a deeper
understanding of polaron physics. In addition, we find that the polaron
spectrum can be used to measure the superfluid pairing gap, and in
the case of a magnetic impurity, the energy of the sub-gap Yu-Shiba-Rusinov
(YSR) bound state. Our theoretical predictions can be examined with
state-of-art cold-atom experiments.
\end{abstract}
\maketitle

\section{Introduction}

The dynamics of an impurity interacting with a bath of quantum-mechanical
particles are unique and fundamental in understanding many-body quantum
physics \citep{Balatsky2006RMP,Bruun2014Review,Schmidt2018Review}.
On the one hand, the system's simplicity allows us to develop insightful
theoretical models and, in some cases, access exact solutions to make
quantitative comparisons with experiments \citep{Nozieres1969PR,Mahan2000Book}.
On the other hand, since a single impurity barely affects the background,
we can apply the impurity as a sensitive probe of the surrounding
many-particle medium \citep{Balatsky2006RMP}. Two important and related
theoretical concepts have been developed to study the impurity-medium
problems: polarons \citep{Bruun2014Review,Schmidt2018Review} and
orthogonality catastrophe (OC) \citep{Anderson1967PRL,Demler2012PRX}.

In 1933, Landau \citep{Landau1933PhysZSoviet} introduced the general
concept of polarons to describe impurity-medium systems as quasiparticles
formed by dressing the impurity with elementary excitations of the
medium. Polarons have become some of the most celebrated ``quasiparticles''
in condensed matter physics and can be commonly found in various crystalline
solids \citep{Chaikin1972PRB,Lindemann1983PRB}. In recent years,
polaron physics in experiment \citep{Schirotzek2009PRL,Zhang2012PRL,Grimm2012Nature,Kohl2012Nature,Demler2016Science,Hu2016PRL,Jorgensen2016PRL,Scazza2017PRL,Yan2019PRL,Zwierlein2020Science,Sagi2020PRX}
and theory \citep{Chevy2006PRA,Lobo2006PRL,Combescot2007PRL,Punk2009PRA,Cui2010PRA,Mathy2011PRL,Schmidt2012PRA,Rath2013PRA,Shashi2014PRA,Li2014PRA,Kroiss2015PRL,Levinsen2015PRL,HuHui2016PRA,Goulko2016PRA,HuHui2018PRA,Mulkerin2019AnnPhys,Jia2019PRL,Isaule2021PRA}
has progressed rapidly in the new platform of ultracold quantum gases,
which provides unprecedented controllability and accessibility \citep{Bloch2008RMP,Chin2010RMP}.
The insightful concept of polaron leads to developing approximate
approaches such as the extended Chevy's ansatz \citep{Chevy2006PRA,Cui2010PRA}
or the many-body $T$-matrix method \citep{Bruun2014Review,Combescot2007PRL,HuHui2018PRA}
that includes only a few medium excitations, which proved to be an
excellent approximation for mobile impurities. The underlying physics
is that multiple medium excitations cost the mobile impurity's recoil
energy and are energetically unfavorable. Together with Monte Carlo
simulations, these approximated approaches predict several characteristic
features of the polaron spectrum: attractive and repulsive polaron
branches with finite residue, the dark continuum \citep{Goulko2016PRA},
and the molecule-hole continuum \citep{Bruun2014Review}. While both
attractive and repulsive polarons have been observed in experiments
\citep{Grimm2012Nature,Kohl2012Nature}, other features remain elusive
due to the uncertainty in theoretical calculations.

In contrast to a mobile impurity, an infinitely heavy impurity immersed
in a Fermi sea can excite many particle-hole pairs close to the Fermi
surfaces without costing recoil energy, leading to the occurrence
of OC \citep{Demler2012PRX,Schmidt2018Review}. The concept of OC,
i.e., the many-particle states with and without impurity become orthogonal,
was raised by Anderson in 1967 \citep{Anderson1967PRL} to understand
the Fermi-edge singularity of x-ray absorption spectra in metals \citep{Nozieres1969PR,Mahan2000Book}.
This well-known Fermi-edge singularity is the first and most important
example of non-equilibrium many-body physics and is exactly solvable
\citep{Weiss1999Book,Rosch1999AdvPhys} via the functional determinant
approach (FDA) \citep{Leonid1996JMathPhys,Klich2003Book,Schonhammer2007PRB,Ivanov2013JMathPhys}.
Unfortunately, OC leads to vanishing quasiparticle residues \citep{Schmidt2018Review},
where polaron does not technically exist. Consequently, this exactly
solvable model may not be directly applied to understand polarons.

The present study, which accompanies the Letter Ref. \citep{AccompanyingShort2022PRL},
investigates a heavy impurity immersed in a two-component Fermi superfluid
medium described by the standard Bardeen-Cooper-Schrieffer (BCS) pairing
theory. The purpose is twofold. 

First, we aim to construct an exactly solvable model for polaron with
finite residue. As shown in this study, our system is suitable for
an exact approach --- an extended FDA, and the presence of a pairing
gap can efficiently suppress multiple particle-hole excitations and
prevent Anderson's OC. Therefore, our model provides a benchmark calculation
of the polaron spectrum and rigorously examine all the speculated
polaron features. We name our system ``heavy crossover polaron''
since the background Fermi gas can undergo a crossover from a Bose-Einstein
condensation (BEC) to a BCS superfluid. 

Second, our prediction can be applied to investigate the background
Fermi superfluid excitations, a long-standing topic in ultracold atoms.
Polarons have already been realized in BEC and ideal Fermi gas experimentally,
but the physics of these weakly interacting background gas are well
understood. More recently, it has also been shown that polarons in
BEC with a synthetic spin-orbit-coupling can reveal the nature of
the background roton excitations \citep{Jia2019PRL}. Investigating
polaron physics in a strongly correlated Fermi superfluid at the BEC-BCS
crossover, namely crossover polaron, has also been proposed in several
pioneering works with approximated approaches \citep{Nishida2015PRL,Yi2015PRA,Pierce2019PRL,HuHui2021arXiv,Bigue2022arXiv}.
Our exact method in the heavy impurity limit allows us to apply the
polaron spectrum to measure the Fermi superfluid excitation features
such as the paring gap and sub-gap Yu-Shiba-Rusinov (YSR) bound states
\citep{Yu1965ActaPhysSin,Shiba1968ProgTheorPhys,Rusinov1969JETP,Vernier2011PRA,Jiang2011PRA},
which is highly experimentally relevant. Nowadays, it is standard
to use Feshbach resonance at the BEC-BCS crossover to realize a BCS
Fermi superfluid. Recent experiments have already demonstrated the
coexistence of Bose and Fermi superfluids in several realizable systems,
$^{6}$Li-$^{7}$Li \citep{Salomon2014Science}, $^{6}$Li-$^{41}$K
\citep{YaoXingCan2016PRL}, and $^{6}$Li-$^{174}$Yb \citep{Roy2017PRL}
mixtures, where the heavy species can serve as the impurity at will.
The combinations $^{6}$Li-$^{133}$Cs \citep{Tung2013PRA}, $^{6}$Li-$^{168}$Er
\citep{Schafer2022PRA}, and $^{6}$Li-$^{168}$Er \citep{Schafer2022PRA}
are also promising candidates, where the interspecies Feshbach resonances
have been characterized.

The rest of this paper is organized as follows. In the following section,
we establish our general formalism and show how to extend the exact
FDA approach to the case of a BCS superfluid as a background system.
Section III is devoted to presenting our numerical results, and Section
IV is given to the discussion of possible experimental realizations.
Finally, we conclude our paper by discussing the physics and proposing
applications in section V. 

\section{Formalism}

\subsection{Heavy impurity in a BCS superfluid}

Our system consists of a static impurity atom, that either is localized
by a deep optical lattice or has infinitely heavy mass, and a two-component
Fermi superfluid with equal mass $m_{\uparrow}=m_{\downarrow}=m$.
We assume the impurity can be in either a non-interacting or an interacting
hyperfine state with the background fermions, where the many-body
Hamiltonian is given by $\hat{H}_{i}$ and $\hat{H}_{f}$, correspondingly.
The energy difference of these two hyperfine states only leads to
trivial effects and is neglected in this work. The interaction between
unlike atoms in the two-component Fermi gas can be tuned by a broad
Feshbach resonance, and characterized by the $s$-wave scattering
length $a$. At low temperature $T$, these strongly interacting fermions
undergo a crossover from a BEC to a BCS superfluid, which can be described
by the celebrated BCS theory at a mean-field level and is briefly
reviewed here and in Appendix \ref{sec:AppA_BCS}.

Using the units $\hbar=1$ hereafter, the BCS Hamiltonian is given
by
\begin{equation}
\hat{\mathcal{H}}_{i}=K_{0}+\sum_{\mathbf{k}}\hat{\psi}_{\mathbf{k}}^{\dagger}\underline{h_{i}(\mathbf{k})}\hat{\psi}_{\mathbf{k}},\label{eq:Hiblinear}
\end{equation}
where $\hat{\psi}_{\mathbf{k}}^{\dagger}\equiv(c_{\mathbf{k}\uparrow}^{\dagger},c_{-\mathbf{k}\downarrow})$
is the Nambu spinor representation with $c_{\mathbf{k}\sigma}^{\dagger}$
($c_{\mathbf{k}\sigma}$) being the creation (annihilation) operator
for a $\sigma$-component fermion with momentum $\mathbf{k}$. Here,
$K_{0}\equiv-\mathcal{V}\Delta^{2}/g+\sum_{\mathbf{k}}(\epsilon_{\mathbf{k}}-\mu)$
with $\mathcal{V}$ denoting the system volume and $\Delta$ being
the pairing gap parameter. $\epsilon_{\mathbf{k}}\equiv\hbar^{2}k^{2}/2m$
is the single-particle dispersion relation and $\mu$ is the chemical
potential. The bare coupling constant $g$ should be renormalized
by the $s$-wave scattering length $a$ between the two components
via 
\begin{equation}
g^{-1}=\frac{m}{4\pi\hbar^{2}a}-\sum_{\mathbf{k}}^{\Lambda}\frac{1}{2\epsilon_{\mathbf{k}}},
\end{equation}
where $\Lambda$ is an ultraviolet cut--off. $\underline{h_{i}(\mathbf{k})}$
can be regarded as a single-particle Hamiltonian $\hat{h}_{i}$ in
momentum space and has a matrix form:

\begin{equation}
\underline{h_{i}(\mathbf{k})}=\left[\begin{array}{cc}
\xi_{\mathbf{k}} & \Delta\\
\Delta & -\xi_{\mathbf{k}}
\end{array}\right],\label{eq:hBCS}
\end{equation}
where $\xi_{\mathbf{k}}\equiv\epsilon_{\mathbf{k}}-\mu$. For a given
scattering length $a$ and temperature $T$, $\Delta$ and $\mu$
are determined by a set of the mean-field number and gap equations
(see Appendix \ref{sec:AppA_BCS}).

When the static impurity is in the interacting hyperfine state, the
many-particle Hamiltonian is given by,

\begin{equation}
\hat{\mathcal{H}}_{f}=\hat{H}_{i}+\hat{V}\equiv\hat{H}_{i}+\sum_{\sigma}\tilde{V}_{\sigma}(\mathbf{k}-\mathbf{q})c_{\mathbf{k}\sigma}^{\dagger}c_{\mathbf{q}\sigma},
\end{equation}
where $\tilde{V}_{\sigma}(\mathbf{k})$ is Fourier transformation
of $V_{\sigma}(\mathbf{r})$, the potential between impurity and $\sigma$-component
fermion. For a reason which will become clear soon, we would like
to express $\hat{H}_{i}$ and $\hat{H}_{f}$ in a bilinear form. Defining
$\hat{\psi}_{\mathbf{k}}^{\dagger}=(c_{\mathbf{k}\uparrow}^{\dagger},c_{-\mathbf{k}\downarrow})\equiv(c_{\mathbf{k}}^{\dagger},h_{\mathbf{k}}^{\dagger})$
and rewriting $\hat{V}$ as
\begin{equation}
\hat{V}=\sum_{\mathbf{k\mathbf{q}}}\left[\tilde{V}_{\uparrow}(\mathbf{k}-\mathbf{q})c_{\mathbf{k}}^{\dagger}c_{\mathbf{q}}-\tilde{V}_{\downarrow}(\mathbf{q}-\mathbf{k})h_{\mathbf{k}}^{\dagger}h_{\mathbf{q}}\right]+\sum_{\mathbf{k}}\tilde{V}_{\downarrow}(0),
\end{equation}
make the bilinear form apparent. We can write the bilinear form of
$\hat{\mathcal{H}}_{f}$ explicitly,

\begin{equation}
\hat{\mathcal{H}}_{f}=K_{0}+\omega_{0}+\sum_{\mathbf{k\mathbf{q}}}\hat{\psi}_{\mathbf{k}}^{\dagger}\underline{h_{f}(\mathbf{k},\mathbf{q})}\hat{\psi}_{\mathbf{q}},\label{eq:Hfbilinear}
\end{equation}
where $\omega_{0}=\sum_{\mathbf{k}}\tilde{V}_{\downarrow}(0)$ and

\begin{equation}
\underline{h_{f}(\mathbf{k},\mathbf{q})}=\underline{h_{i}(\mathbf{k})}\delta_{\mathbf{k}\mathbf{q}}+\left[\begin{array}{cc}
\tilde{V}_{\uparrow}(\mathbf{k}-\mathbf{q}) & 0\\
0 & -\tilde{V}_{\downarrow}(\mathbf{q}-\mathbf{k})
\end{array}\right]
\end{equation}
can be regarded as a single-particle Hamiltonian $\hat{h}_{f}$ in
momentum space and in a matrix form. We can see that, $\hat{h}_{i}$
and $\hat{h}_{f}$ are the single-particle representative of $\hat{\mathcal{H}}_{i}$
and $\hat{\mathcal{H}}_{f}$ up to some constants, respectively.

\begin{figure*}
\includegraphics[width=0.9\textwidth]{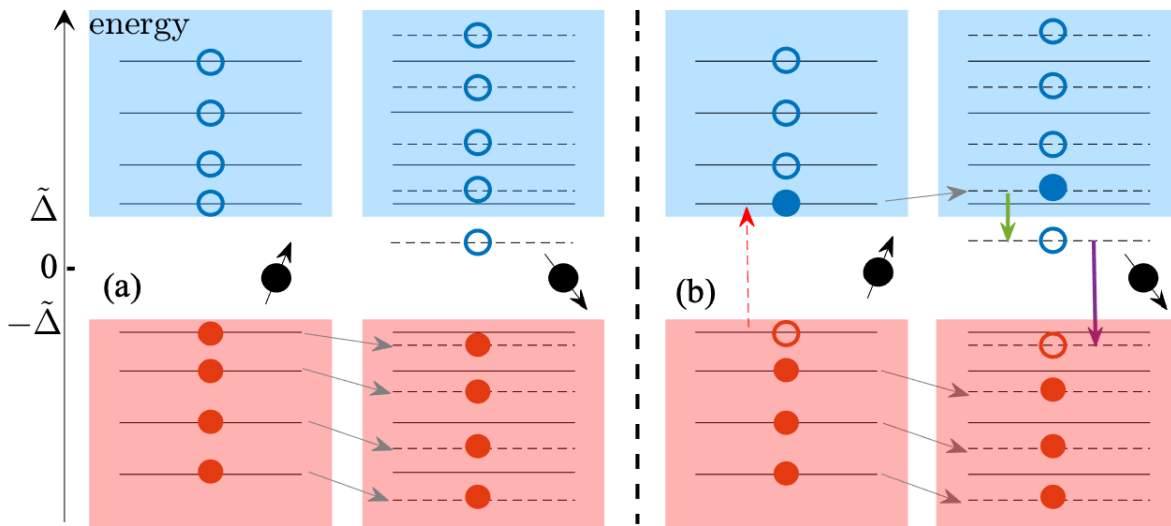}\caption{A sketch of the occupation and structure of the single-particle spectrum
of a two-component superfluid Fermi gas at (a) zero temperature and
(b) finite temperature. The big black sphere represents the impurity
in non-interacting (black arrow up) or interacting (black arrow down)
hyperfine state. These spectra always have two-branch structures separated
by an energy gap of $2\tilde{\Delta}$, where the red and blue rectangles
represent the negative and blue branches, respectively. However, the
individual energy level with or without impurity interaction have
shifts as indicated by the black solid and dashed lines, correspondingly.
There also exists an in-gap YSR bound state if the impurity interaction
is magnetic ($a_{\uparrow}\protect\ne a_{\downarrow}$).\label{fig:SpectrumSketch}}
\end{figure*}

It is worth noting that, in the many-body Hamiltonian $\hat{\mathcal{H}}_{f}$
we have assumed that the pairing order parameter $\Delta$ remains
unchanged by introducing the interaction potential $V_{\sigma}(\mathbf{r})$.
For a non-magnetic potential ($V_{\uparrow}=V_{\downarrow}$) that
respects time-reversal symmetry, this is a reasonable assumption,
according to Anderson's theorem \citep{Balatsky2006RMP}. For a magnetic
potential ($V_{\uparrow}\neq V_{\downarrow}$), the local pairing
gap near the impurity will be affected, as indicated by the presence
of the YSR bound state. We will follow the typical non-self-consistent
treatment of the magnetic potential in condensed matter physics \citep{Balatsky2006RMP,Yu1965ActaPhysSin}
and assume a constant pairing gap as the first approximation for simplicity.
We leave a more rigorous self-consistent calculation of a pairing
gap to future studies.

\subsection{Functional determinant approach}

We are interested in a situation where the impurity is driven from
a non-interacting hyperfine state to an interacting hyperfine state
at $t=0$, as sketched in Fig. \ref{fig:SpectrumSketch}. The most
basic quantity to describe the response to this process is the time-overlapping
function
\begin{equation}
S(t)=\left\langle \mathrm{e}^{\mathrm{i}\hat{\mathcal{H}}_{i}t}\mathrm{e}^{-\mathrm{i}\hat{\mathcal{H}}_{f}t}\right\rangle \equiv{\rm Tr}\left[\mathrm{e}^{\mathrm{i}\hat{\mathcal{H}}_{i}t}\mathrm{e}^{-\mathrm{i}\hat{\mathcal{H}}_{f}t}\hat{\rho}_{\mathrm{0}}\right],\label{eq:St}
\end{equation}
where $\hat{\rho}_{0}$ is the initial thermal density matrix, and
$\hat{H}_{i}$ and $\hat{H}_{f}$ are the many-body Hamiltonian with
the impurity in non-interacting and interacting hyperfine state, respectively.
The response in frequency domain can be obtained via a Fourier transformation
\begin{equation}
A(\omega)=\frac{1}{\pi}{\rm Re}\int_{0}^{\infty}e^{i\omega t}S(t)dt,
\end{equation}
which is also called spectral function.

We review our main theoretical tool, FDA, and show how to extend this
method to the case of an ultracold BCS superfluid as the background
medium. An exact calculation of Eq. (\ref{eq:St}) is usually not
accessible due to the exponentially growing complexity of the many-body
Hamiltonian with respect to particle number $N$. However, one can
prove that Eq. (\ref{eq:St}) can reduce to a determinant in a single-particle
Hilbert space that grows only linearly to $N$, if $\hat{H}_{i}$
and $\hat{H}_{f}$ are both fermionic, bilinear many-body operators,
such as Eq. (\ref{eq:Hiblinear}) and Eq. (\ref{eq:Hfbilinear}) shown
in the previous subsection. In that case, we have,
\begin{equation}
S(t)=e^{-i\omega_{0}t}{\rm det}[1-\hat{n}+e^{i\hat{h}_{i}t/\hbar}e^{-i\hat{h}_{f}t/\hbar}\hat{n}]\label{eq:FDA}
\end{equation}
where $\hat{n}$ is the occupation number operator.

It would be more convenient to carry out numerical calculations in
the coordinate space in a finite system confined in a sphere of radius
$R$. We then take the system size towards infinity, while keeping
the density constant, until numerical results are converged. The bilinear
form of the many-body Hamiltonians in coordinate space are given by
\begin{equation}
\hat{\mathcal{H}}_{i}=K_{0}+\int d\mathbf{r}\hat{\phi}^{\dagger}(\mathbf{r})\underline{h_{i}(\mathbf{r})}\hat{\phi}(\mathbf{r}),
\end{equation}
\begin{equation}
\hat{\mathcal{H}}_{f}=K_{0}+\omega_{0}+\int d\mathbf{r}\hat{\phi}^{\dagger}(\mathbf{r})\underline{h_{f}(\mathbf{r})}\hat{\phi}(\mathbf{r}),
\end{equation}
with $K_{0}=-\mathcal{V}\Delta^{2}/g+\sum_{\mathbf{k}}(\epsilon_{\mathbf{k}}-\mu)$
being an unimportant constant that cancels out in Eq. (\ref{eq:FDA}).
Here, $\hat{\phi}^{\dagger}(\mathbf{r})=[c_{\uparrow}^{\dagger}(\mathbf{r}),c_{\downarrow}(\mathbf{r})]\equiv[c^{\dagger}(\mathbf{r}),h^{\dagger}(\mathbf{r})]$
are creation operators in the coordinate space. Since higher partial
wave interaction is negligible at low temperature, we focus on the
$s$-wave channel. We also assume $V_{\sigma}(\mathbf{r})=V_{\sigma}(r)$
is spherically symmetric and short-range. The single-particle representative
of Hamiltonians in coordinate space, therefore, are given by

\begin{equation}
\begin{aligned}\underline{h_{f}(r)}= & \underline{h_{i}(r)}+\left(\begin{array}{cc}
V_{\uparrow}(r) & 0\\
0 & -V_{\downarrow}(r)
\end{array}\right)\\
\equiv & \left(\begin{array}{cc}
-\frac{1}{2m}\frac{d^{2}}{dr^{2}}+V_{\uparrow}(r)-\mu & \Delta\\
\Delta & \frac{1}{2m}\frac{d^{2}}{dr^{2}}-V_{\downarrow}(r)+\mu
\end{array}\right).
\end{aligned}
\end{equation}
In our numerical calculation, we choose a soft-core van-der-Waals
potential 
\begin{equation}
V_{\sigma}(r)=-\frac{C_{6}}{r^{6}}\exp\left[-\frac{r_{\sigma}^{6}}{r^{6}}\right],
\end{equation}
where $C_{6}$ is the dispersion coefficient describing the long-range
behavior of the impurity-fermion interaction and determines the van-der-Waals
length $l_{{\rm vdW}}=(2mC_{6})^{1/4}/2$. The short-range parameter
$r_{\sigma}$ are tuned to give the desired energy-dependent scattering
length 
\begin{equation}
a_{\sigma}(E_{F})=-\frac{\tan\eta_{\sigma}(k_{F})}{k_{F}},
\end{equation}
where $\eta_{\sigma}(k_{F})$ is the $s$-wave scattering length between
the impurity and $\sigma$-component fermions at the Fermi energy
$E_{F}=k_{F}^{2}/2m$. We find our calculations are insensitive to
other details of $V_{\sigma}(r)$ (such as the value of $l_{{\rm vdW}}$
and the number of short-range bound states the potential supported)
as long as $k_{F}l_{{\rm vdW}}\ll1$. Therefore, we denote $a_{\sigma}(E_{F})\equiv a_{\sigma}$
hereafter for the simplicity of notation. In the calculations here,
we choose $k_{F}l_{{\rm vdW}}=0.01$ unless specify otherwise. To
calculate Eq. (\ref{eq:FDA}), we need to find the eigenpairs $E_{\nu}$,
$\phi_{\nu}\equiv[\phi_{\nu,\uparrow}(r),\phi_{\nu,\downarrow}(r)]$
for $\underline{h_{i}(r)}$ and $\tilde{E}_{\nu}$, $\tilde{\phi}_{\nu}$
for $\underline{h_{f}(r)}$, and express the occupation operator $\hat{n}$
as a diagonal matrix with elements 
\begin{equation}
n_{\nu\nu}=f(E_{\nu})=\frac{1}{e^{-E_{\nu}/k_{B}T}+1}.
\end{equation}
We also need to take care of $\omega_{0}$ if $V_{\downarrow}\ne0$.
Noticing that the original definition $\omega_{0}=\sum_{\mathbf{k}}\tilde{V}_{\downarrow}(0)=\sum_{\mathbf{k}}\langle\mathbf{k}|\mathbf{r}\rangle V(\mathbf{r})\langle\mathbf{r}|\mathbf{k}\rangle$
is equivalent to taking the trace of the matrix representative of
$\hat{V}_{\downarrow}$ in momentum state basis. Therefore, $\omega_{0}$
can also be obtained via tracing $\underline{V_{\downarrow}}$, the
matrix format of $\hat{V}_{\downarrow}$ in an arbitrary complete
orthogonal set of basis, i.e., $\omega_{0}={\rm Tr}\underline{V_{\downarrow}}$.
In practical calculations, we use the same discretization basis set
in coordinate space as the one applied to diagonalize $\underline{h_{i}(r)}$
and $\underline{h_{f}(r)}$.

We give a few further remarks on some possible extensions of our methods.
As already noticed in Ref. \citep{Demler2012PRX}, generalization
of FDA to other geometries and confinement can be easily implemented
to the single-particle Hamiltonian. Our single-channel soft-core van-der-Waals
potential have been proved to mimic the interatomic interaction near
broad Feshbach resonances very well \citep{Jia2012PRL}, and can be
replaced by multi-channel interactions to describe closed-channel
dominated Feshbach resonances.

\begin{figure}
\includegraphics[width=0.98\columnwidth]{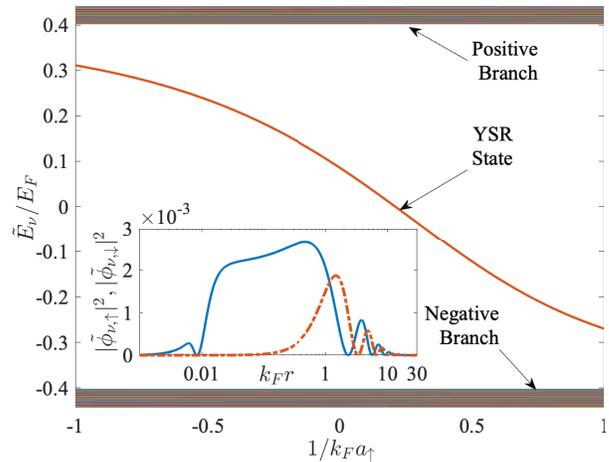}\caption{Single-particle spectrum of a Fermi superfluid with a magnetic impurity
($a_{\downarrow}=0$) as a function of $1/(k_{F}a_{\uparrow})$. The
scattering length between the two-component fermions is $k_{F}a=-2$,
which gives rise to $\mu\simeq0.85E_{F}$ and $\Delta\simeq0.40E_{F}$
at zero temperature. The solid red curve in the middle shows the YSR
bound state energy. The inset shows the corresponding YSR wave-functions
$\tilde{\phi}_{\nu,\uparrow}$ (blue solid curve) and $\tilde{\phi}_{\nu,\downarrow}$
(red dash-dotted curve) at $k_{F}a_{\uparrow}=-2$. \label{fig:YSR}}
\end{figure}

\section{Results}

\subsection{Single particle spectrum}

It is illustrative to first see the structure of single particle spectrum,
as sketched in Fig. \ref{fig:SpectrumSketch}. When the impurity interaction
is absent, diagonalizing $\underline{h_{i}(r)}$ gives the well-known
BCS dispersion relation

\begin{equation}
E_{\nu}=\pm\mathcal{E}_{k_{\nu}}=\pm\sqrt{\xi_{k_{\nu}}^{2}+\Delta^{2}},
\end{equation}
where $k_{\nu}R=n_{\nu}\pi$ with integer $n_{\nu}$. The positive
and negative branches of the spectrum are separated by an energy gap

\begin{equation}
2\tilde{\Delta}=\begin{cases}
2\Delta & \mu\ge0\\
2\sqrt{\Delta^{2}+\mu^{2}} & \mu<0
\end{cases},
\end{equation}
which represents the minimum energy required to break a Cooper-pair
into a particle-hole excitation. At zero temperature, the many-body
ground state can be regarded as a fully filled Fermi sea of the lower
branch, and a completely empty Fermi sea of the upper branch. {[}Notice
that the $E_{\nu}$ are measured with respect to chemical potential
$\mu$, which leads to the occupation $f(E_{\nu})=1/\left(e^{-E_{\nu}/k_{B}T}+1\right)${]}.

In the presence of impurity interaction, our numerical calculations
show that $\tilde{E}_{\nu}$ still consists of two branches separated
by $2\tilde{\Delta}$, with each individual energy level shifted as
shown in Fig. \ref{fig:SpectrumSketch}. Moreover, when the impurity
scattering is magnetic ($a_{\uparrow}\ne a_{\downarrow})$, a sub-gap
YSR bound state exists. Figure \ref{fig:YSR} shows the YSR bound
state energy as a function of $1/(k_{F}a_{\uparrow})$ for the case
$k_{F}a=-2$ and $k_{F}a_{\downarrow}=0$ at zero temperature. The
decreasing bound state energy with increasing $1/(k_{F}a_{\uparrow})$
can be qualitatively understood from the analytic expression 
\begin{equation}
E_{\textrm{YSR}}\simeq\Delta\cos\left[\eta_{\uparrow}(E_{F})-\eta_{\downarrow}(E_{F})\right],\label{eq:EYSR}
\end{equation}
which holds in the weak-coupling limit ($a\rightarrow0^{-}$) \citep{Rusinov1969JETP}.
Here, $\eta_{\uparrow}(E_{F})$ and $\eta_{\downarrow}(E_{F})=0$
are the impurity scattering phase shifts of the potentials $V_{\uparrow}(r)$
and $V_{\downarrow}(r)$ at Fermi energy $E_{F}$. The inset of Fig.
\ref{fig:YSR} shows the YSR wave-function at $k_{F}a_{\uparrow}=-2$,
where one can see that the YSR bound state has a relatively large
size (about 30$k_{F}^{-1}$ in this case) and shows an oscillation
behavior at large distances.

We give some further remarks here on the two-body bound states supported
solely by the short-range potential $V_{\sigma}(r)$, when the other
component of fermions are absent. In general, there are multiple such
bound states, and almost all of them are deeply bound with large binding
energy $E_{b}\gg\Delta$ and highly localized to the impurity. As
a result, the overlapping between these deeply bound states and BCS
scattering waves are vanishingly small. Therefore, these deeply bound
states are almost unaffected by the presence of the other component
and give negligible effects on the response functions. The only exception
is the shallowest bound state with $a_{\sigma}>0$. This shallow bound
state can strongly couple to the scattering states of the other component,
and hence can no longer be distinguished from the eigenstates $\tilde{\phi}_{\nu}$.

\begin{figure}
\includegraphics[width=0.98\columnwidth]{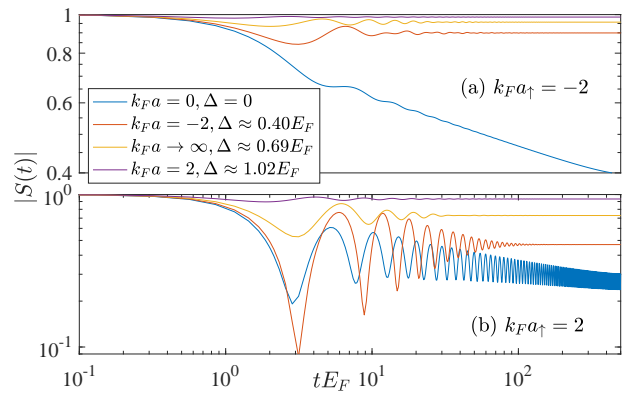}\caption{Zero-temperature Ramsey responses $|S(t)|$ for a magnetic impurity
($a_{\downarrow}=0$) scattering with (a) attractive scattering lengths
$a_{\uparrow}<0$ and (b) repulsive scattering lengths $a_{\uparrow}>0$
are shown for different values of the scattering length $a$ between
the two-component fermions; see legend. \label{fig:StMag}}
\end{figure}

\begin{figure}
\includegraphics[width=0.98\columnwidth]{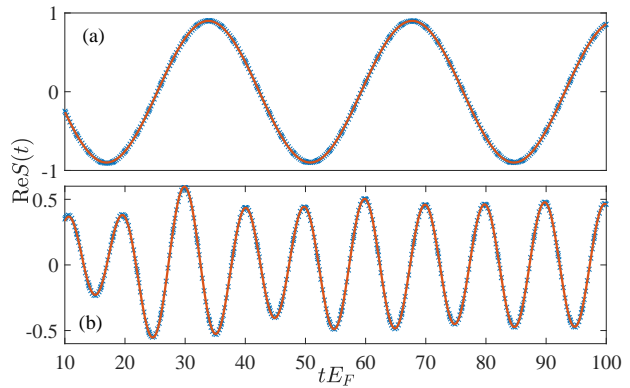}\caption{Fitting asymptotic behavior of ${\rm Re}S(t)$ at zero temperature
for a magnetic impurity ($a_{\downarrow}=0$) that only interacts
with spin-up component of the Fermi superfluid with $k_{F}a=-2$ and
(a) $k_{F}a_{\uparrow}=-2$ (b) $k_{F}a_{\downarrow}=2$. The cross
symbols are the numerical results, and the solid curves correspond
to the fitting formula Eq. (\ref{eq:Sfit_Mag}). \label{fig:MagStFit}}
\end{figure}

\subsection{Magnetic impurity}

We first focus on the simplest case, where the impurity only interacts
with the spin-up component, i.e. $a_{\downarrow}=0$. 

When $\Delta=0$ and $a_{\downarrow}=0$, our system reduces back
to an ideal Fermi gas (consisting of spin-up fermions), and the asymptotic
behavior of the Ramsey response at $t\rightarrow\infty$ is given
by
\begin{equation}
\begin{aligned}S(t)\simeq & Ce^{-i\Delta Et/\hbar}\left(\frac{1}{iE_{F}t/\hbar+0^{+}}\right)^{\alpha}\\
 & +C_{b}e^{-i\left(\Delta E-E_{F}+E_{b}\right)t/\hbar}\left(\frac{1}{iE_{F}t/\hbar+0^{+}}\right)^{\alpha_{b}},
\end{aligned}
\label{eq:Sfit_idealF}
\end{equation}
where $C$ and $C_{b}$ are both numerical constants independent with
respect to $k_{F}a$ and $C_{b}=0$ for $a_{\uparrow}<0$. Here, 
\begin{equation}
\alpha=\eta_{\uparrow}(E_{F})^{2}/\pi^{2}
\end{equation}
and 
\begin{equation}
\alpha_{b}=[1+\eta_{\uparrow}(E_{F})/\pi]^{2}
\end{equation}
are determined by the scattering phase shifts $\eta_{\uparrow}(E_{F})$
at Fermi energy. $E_{b}$ is the binding energy of the shallowest
bound state consisting of the impurity and a spin-up fermion for $a_{\uparrow}>0$
and $\Delta=0$. Furthermore, the change in energy is given by
\begin{equation}
\Delta E=\sum_{E_{\nu}<0}(E_{\nu}-\tilde{E}_{\nu}),
\end{equation}
where deeply bound states are excluded from $\tilde{E}_{\nu}$. Notice
that the power-law decaying behavior of $|S(t)|$ at $\Delta=0$ is
evident in Fig. \ref{fig:StMag} (see the blue lines).

\begin{figure*}
\includegraphics[width=0.7\textwidth]{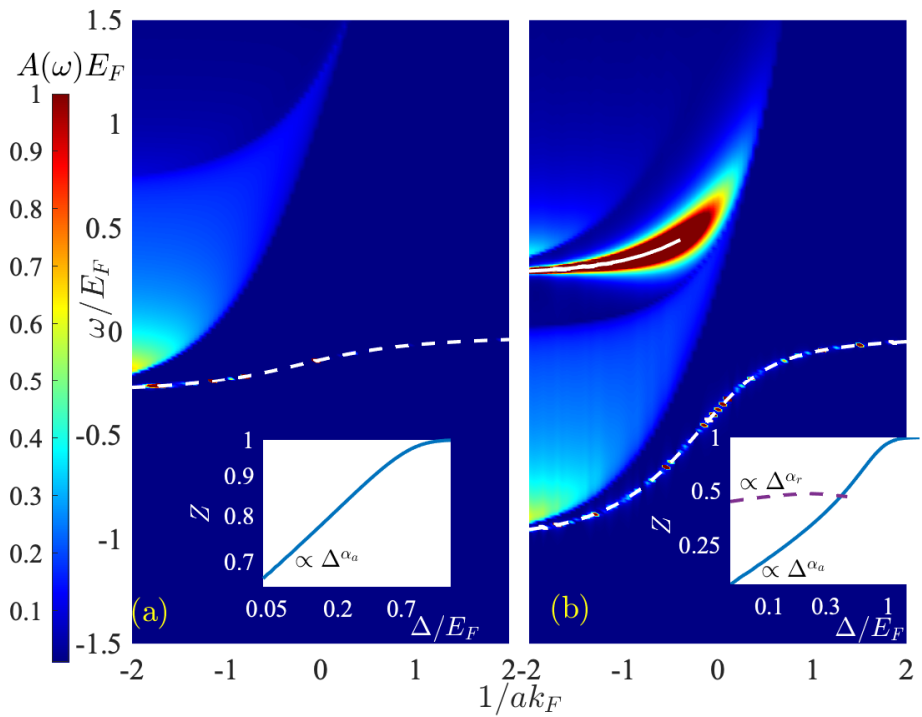}

\caption{Zero-temperature polaron spectra with $a_{\downarrow}=0$ for (a)
$k_{F}a_{\uparrow}=-2$ and (b) $k_{F}a_{\uparrow}=2$ as a function
of $1/(k_{F}a)$ at the BEC-BCS crossover. The white dashed and solid
curves corresponds to the attractive ($E_{a}$) and repulsive ($E_{r}$)
polaron energy, respectively. The insets shows the residues of the
polarons. The blue solid curves and the purple dashed curves show
the residue of attractive polaron $Z_{a}$ and repulsive polaron $Z_{r}$,
respectively, as a function of $\Delta$, which show power-law behaviors
at small $\Delta$. \label{fig:MagPol}}
\end{figure*}

In shark contrast, for cases with nonzero pairing gap, the asymptotic
behavior in the long-time limit shows that $|S(t\rightarrow\infty)|\propto t^{0}$
approach to some constants. These asymptotic constants are larger
for larger $\Delta$. Further details can be obtained by an asymptotic
form that fits our numerical calculations perfectly well, as reported
in Fig. \ref{fig:MagStFit},
\begin{equation}
S(t)\simeq D_{a}e^{-iE_{a}t}+D_{r}e^{-iE_{r}t},\label{eq:Sfit_Mag}
\end{equation}
where $D_{r}=0$ for $a_{\uparrow}<0$. We obtain $D_{a}$, $D_{r}$,
$E_{a}$ and $E_{r}$ from fitting, and find that $E_{r}={\rm Re}E_{r}+i{\rm Im}E_{r}$
is in general complex. In contrast, $E_{a}=\sum_{E_{\nu}<0}(E_{\nu}-\tilde{E}_{\nu})$
(where $\tilde{E}_{\nu}$ excludes the two-body deeply bound states)
is purely real, and can be explained as a renormalization of the filled
Fermi sea, as indicated by the grey arrows in Fig. \ref{fig:SpectrumSketch}(a).

The long-time asymptotic behavior of $S(t)$ manifests itself as some
characterized lineshape in the spectral function 
\begin{equation}
A(\omega)\propto\begin{cases}
Z_{a}\delta(\omega-E_{a}) & \omega\approx E_{a}\\
Z_{r}\frac{\left|{\rm Im}E_{r}\right|/\pi}{(\omega-{\rm Re}E_{r})^{2}+({\rm Im}E_{r})^{2}} & \omega\approx{\rm Re}E_{r}
\end{cases},
\end{equation}
i.e., a $\delta$-function around $E_{a}$ and a Lorenzian around
${\rm Re}E_{r}$. The existence of $\delta$-function peak unambiguously
confirms the existence of a well-defined quasiparticle -- the attractive
polaron with energy $E_{a}$. The Lorenzian, on the other hand, can
be recognized as a repulsive polaron with finite width and hence finite
life-time. Here, $Z_{a}=|D_{a}|$ and $Z_{r}=|D_{r}|$ are the residue
of attractive and repulsive polaron, correspondingly. Numerically,
we find that $Z_{a}\propto(\Delta/E_{F})^{\alpha_{a}}$ and $Z_{r}\propto(\Delta/E_{F})^{\alpha_{r}}$
at small $\Delta$ as shown in the insets of Figs. \ref{fig:MagPol}(a)
and \ref{fig:MagPol}(b). As a result, Eq. (\ref{eq:Sfit_Mag}) have
the same form as Eq. (\ref{eq:Sfit_idealF}), the analytic expression
of $S(t)$ for a non-interacting Fermi gas medium, if we replace the
low-energy cut-off $1/t\rightarrow\Delta$. However, the power-law
coefficients $\alpha_{a}$ and $\alpha_{r}$ are only close to but
not exactly the same as the analytical expressions of $\alpha$ and
$\alpha_{b}$. In the inset of Fig. \ref{fig:MagPol}(a), our numerical
fitting gives $\alpha_{a}\approx0.136$, comparing with $\alpha\approx0.124$
for ideal Fermi gases. In the inset of Fig. \ref{fig:MagPol}(b),
$\alpha_{r}\approx0.083$ and $\alpha_{a}\approx0.452$, in compare
with $\alpha\approx0.124$ and $\alpha_{b}\approx0.419$. These small
differences are probably due to the modification of scattering phase-shifts
in the presence of $\Delta$.

\begin{figure*}
\includegraphics[width=1\textwidth]{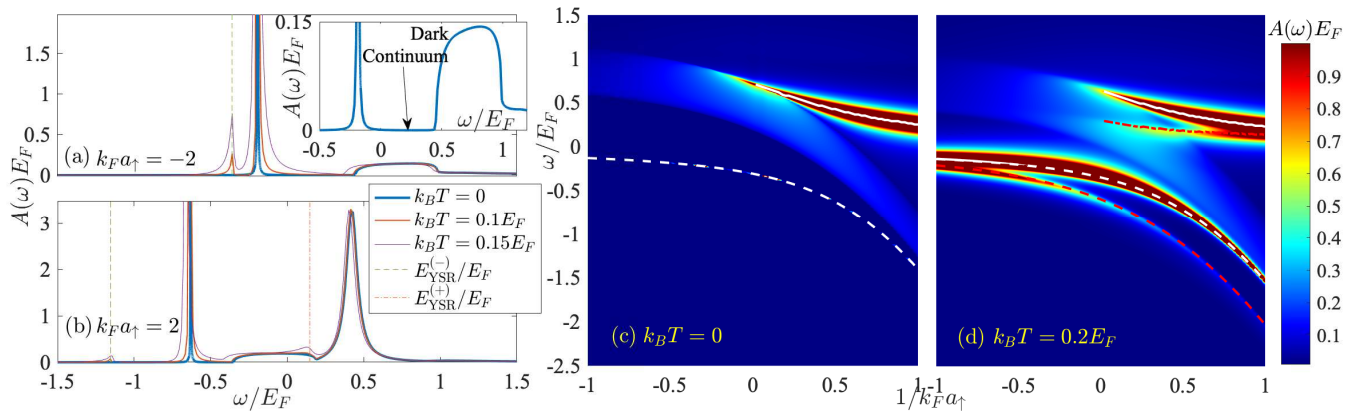}\caption{Polaron spectra with $k_{F}a=-2$ and $a_{\downarrow}=0$ at different
temperature (see legend) with (a) $k_{F}a_{\uparrow}=-2$ and (b)
$k_{F}a_{\uparrow}=2$, with the green dashed and red dash-dotted
vertical lines indicates the YSR features $E_{{\rm YSR}}^{(-)}$ and
$E_{{\rm YSR}}^{(+)}$, respectively. A zoom-in of the dark spectrum
at zero temperature is shown in the inset. A small artificial width
is added to the $\delta$-function peak at zero temperature for visibility.
The full spectra as a function of $1/(k_{F}a_{\uparrow})$ are shown
in (c) at zero temperature and (d) $k_{B}T=0.2$. The white dahsed
and solid curves shows the attractive and repulsive polaron energies,
respectively. The red dashed and dash-dotted in (d) corresponds to
$E_{{\rm YSR}}^{(-)}$ and $E_{{\rm YSR}}^{(+)}$, respectively. \label{fig:AwM}}
\end{figure*}

Next, we study the full zero-temperature polaron spectrum across the
BEC-BCS crossover and show them in Fig. \ref{fig:MagPol}. Numerically,
to obtain $A(\omega)$ accurately requires a Fourier transformation
that involves an integration of $S(t)$ from $t=0$ to $t\rightarrow\infty$.
We follow the procedures adopted from Ref. \citep{Demler2012PRX}:
we numerically integrate $S(t)$ up to some large cut-off time $t^{*}\sim500/E_{F}$,
and carry-out the integration analytically with the fitting formula
Eq. (\ref{eq:Sfit_Mag}) for $t>t^{*}$.

Figure \ref{fig:MagPol}(a) shows the case $k_{F}a_{\uparrow}=-2>0$,
where the white dashed curve indicates the attractive polaron $\delta$-function
peak. This attractive polaron separates from a molecule-hole continuum
by a region of anomalously low spectral weight, namely the ``dark
continuum'' (also shown in the inset of Fig. \ref{fig:AwM}). The
existence of dark continuum has been previously observed in spectra
of other polaron systems. However, most of these studies apply various
approximations, and only recently a diagrammatic Monte Carlo study
proves the dark continuum is indeed physical \citep{Goulko2016PRA}.
Here, our FDA calculation of the heavy crossover polaron spectrum
gives an exact proof of the dark continuum. In addition, we can see
that the dark continuum regime becomes smaller towards the deep BCS
side of the Feshbach resonance for the background Fermi superfluid.
We expect that the dark continuum vanishes in the $\Delta\rightarrow0$
limit, and the attractive polaron will merge into the molecule-hole
continuum, forming a power-law singularity seen in the spectrum of
heavy impurity in an ideal Fermi gas \citep{Demler2012PRX}. 

The white solid curve in Fig. \ref{fig:MagPol}(b) shows the repulsive
polaron energy. We can observe that the repulsive polaron width become
larger from the BCS side towards the unitary limit. Near the unitary
limit, the repulsive polaron residue $Z_{r}$ also deviates from the
power law-dependence and starts decreasing as shown in the inset of
Fig. \ref{fig:MagPol}(b). Towards the BEC side, both the repulsive
polaron and the molecule-hole continuum are vanishing, which can also
be inferred from the behavior $Z_{a}\rightarrow1$ on the deep BEC
side.

We also study the finite-temperature spectrum at $k_{F}a=-2$ as shown
in Fig. \ref{fig:AwM}. Figures \ref{fig:AwM}(a) and \ref{fig:AwM}(b)
show the spectrum at $k_{F}a_{\uparrow}=-2$ and $k_{F}a_{\uparrow}=2$,
respectively. As temperature increases, we observe the expected thermal
broadening and slightly shifts of the spectral peaks since $\Delta$
reduces at finite temperature. Interestingly, we also observe some
additional features. An onset of spectral weight enhancement arises
sharply at the energy
\begin{equation}
E_{{\rm YSR}}^{(-)}=E_{a}-(\Delta-E_{{\rm YSR}}),\label{eq:EYSR-}
\end{equation}
below the attractive polaron. We explain this feature as an additional
decay from the upper branch state to the sub-gap YSR state illustrated
by the green arrow in Fig. \ref{fig:SpectrumSketch}(b). There is
also another feature that associates with the repulsive polaron shows
up for the $k_{F}a_{\uparrow}=2$ case at energy

\begin{equation}
E_{{\rm YSR}}^{(+)}={\rm Re}(E_{r})-(E_{{\rm YSR}}+\Delta),\label{eq:EYSR+}
\end{equation}
which implies that this feature is related to the decay from the YSR
state back to the lower branch as illustrated by the purple arrow
in Fig. \ref{fig:SpectrumSketch}(b). These two decay processes are
only allowed if the upper branch has thermal occupations initially,
which explain why such features only show up at finite temperature.
These features can be better depicted in the comparison of the full
spectra as a function of $1/(k_{F}a_{\uparrow})$ at zero and finite
temperature in Figs. \ref{fig:SpectrumSketch}(c) and \ref{fig:SpectrumSketch}(d),
respectively.

\begin{figure*}
\includegraphics[width=1\textwidth]{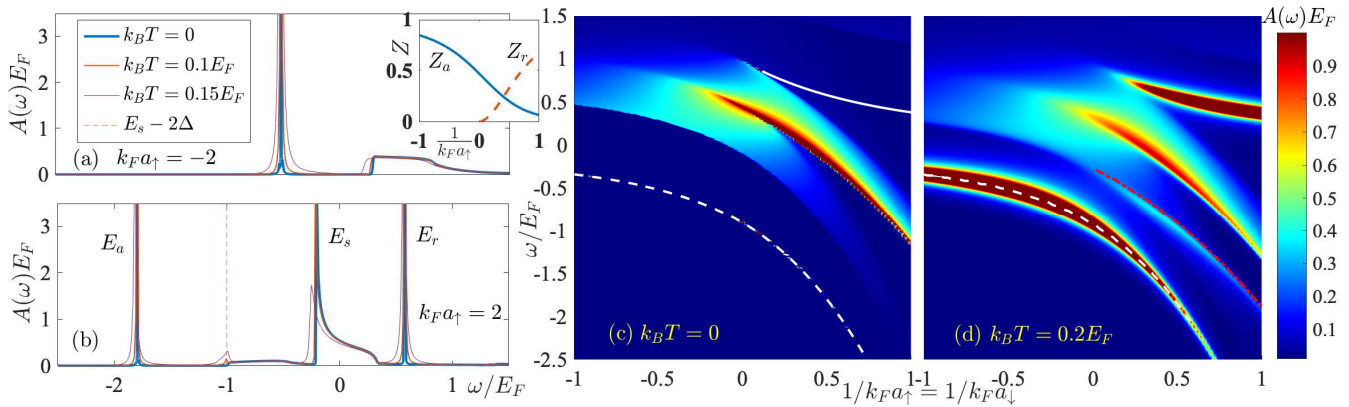}\caption{Polaron spectrum of heavy non-magnetic impurity ($a_{\uparrow}=a_{\downarrow}$)
in a BCS superfluid with $k_{F}a=-2$ at different temperature (see
legend). The impurity scattering length is (a) $k_{F}a=-2$ and (b)
$k_{F}a=2$. The red dashed vertical line shows a feature at $E_{s}-2\Delta$
associated with the singularity at $E_{s}$. The inset shows the residue
as a function of $1/(k_{F}a_{\uparrow})$. The full spectrum as a
function of $1/(k_{F}a_{\uparrow})$ are shown in (c) and (d) for
zero and finite temperature, respectively. The white dashed and solid
curves shows attractive and repulsive polaron energy, and the red
dash-dotted curve shows the finite temperature feature. \label{fig:AwNM}}
\end{figure*}

\begin{figure*}
\includegraphics[width=0.7\textwidth]{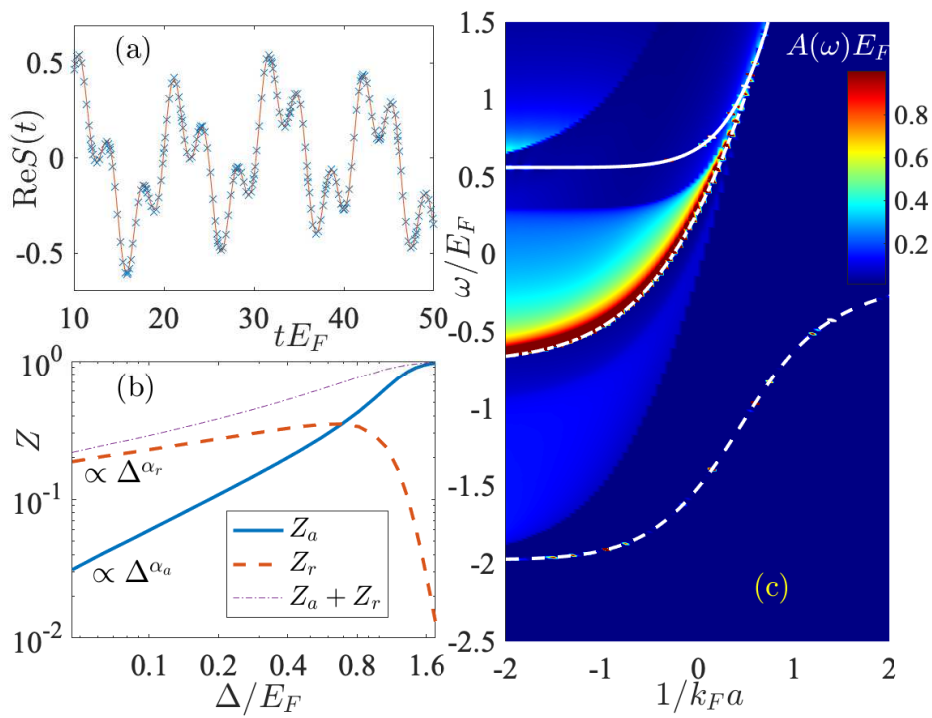}\caption{(a) ${\rm Re}S(t)$ as a function of $t$, the cross symbol shows
the numerical result, and the solid line is the fitting formula Eq.
(\ref{eq:Sfit_NM}). (b) The polaron residue as a function of $\Delta$.
The blue solid, red dashed and purple dash-dotted curves correspond
to $Z_{a}$, $Z_{r}$ and $Z_{a}+Z_{r}$, respectively. The power-law
exponents $\alpha_{a}\approx0.85$ and $\alpha_{r}\approx0.25$. (c)
Polaron spectrum of heavy non-magnetic impurity $(k_{F}a_{\uparrow}=k_{F}a_{\downarrow}=2)$
at zero temperature as a function of $1/(k_{F}a)$ at the BEC-BCS
crossover. The white solid, dash and dash-dotted curve corresponds
to the repulsive polaron, attractive polaron and the singularity energy,
respectively. \label{fig:NMPol}}
\end{figure*}

\subsection{Non-magnetic impurity}

In this subsection, we study the case of non-magnetic impurity scattering
$a_{\uparrow}=a_{\downarrow}$, where the YSR state merges into the
upper branch states as a result of Eq. (\ref{eq:EYSR}) and ceases
to exist. 

As expected, Fig. \ref{fig:AwNM}(a) shows no YSR features at $k_{F}a=-2$
and is quite simple. In contrast, the polaron spectra on the positive
side $k_{F}a_{\uparrow}=2$ are much more complex as depicted in Fig.
\ref{fig:AwNM}(b). Interestingly, the repulsive polaron at zero temperature
is also characterized by a $\delta$-function with infinite life-time.
In addition, another singularity shows up at $E_{s}$. We speculate
the new long-lived repulsive polaron is related to undamped density
excitations (i.e., the gapless Goldstone mode of the Fermi superfluid)
excited by the non-magnetic impurity potential. As the coupling to
the gapless Goldstone mode does not cost energy, the OC mechanism
may lead to a power-law singularity, which is the reminiscent of the
damped repulsive polaron in the case of magnetic impurity scattering.
With this understanding in mind, we have checked that the asymptotic
$t\rightarrow\infty$ behavior fits the formula
\begin{equation}
S(t)\approx D_{a}e^{-iE_{a}t}+D_{r}e^{-iE_{r}t}+D_{s}e^{-iE_{s}t}(\frac{1}{iE_{F}t})^{\alpha_{s}}\label{eq:Sfit_NM}
\end{equation}
very well, as shown in Fig. \ref{fig:NMPol}(a). Our numerical fitting
confirms $E_{a}$, $E_{r}$ and $E_{s}$ are all purely real. We also
find that the power-law component of the singularity $\alpha_{s}\approx0.5$,
which seems to be a constant insensitive to $a_{\uparrow}$, $a_{\downarrow}$
and $a$. The residue $Z_{a}=|D_{a}|$ and $Z_{r}=|D_{r}|$ as a function
of impurity interaction $1/(k_{F}a_{\uparrow})$ are shown in the
inset of Fig. \ref{fig:AwNM}(a), which shows that the attractive
polaron residue decreases and repulsive polaron becomes dominated
on the positive side of impurity scattering length $a_{\uparrow}>0$.
The dependence of $Z_{a}$ and $Z_{r}$ on $\Delta$ is reported in
Fig. \ref{fig:NMPol}(b). Similar to the magnetic impurity case, we
observe the power-law dependences $Z_{a}\propto(\Delta/E_{F})^{\alpha_{a}}$
and $Z_{r}\propto(\Delta/E_{F})^{\alpha_{r}}$ at small $\Delta$,
and $Z_{a}\rightarrow1$ and $Z_{r}\rightarrow0$ on the deep BEC
side $\Delta\rightarrow\infty$. 

Figures \ref{fig:AwNM}(c) and \ref{fig:AwNM}(d) show the comparison
between zero and finite temperature polaron spectrum as a function
of $1/(k_{F}a_{\uparrow})$. We can observe a finite temperature feature
appears at $E_{s}-2\Delta$, as shown in red dash-dotted curve in
Fig. \ref{fig:AwNM}(d) {[}as indicated by the dashed vertical line
in Fig. \ref{fig:AwNM}(b) at $k_{F}a_{\uparrow}=2${]}. This feature
is the reminiscent of the structure at $E_{\textrm{YSR}}^{(+)}$ in
the case of magnetic impurity scattering (see Eq. (\ref{eq:EYSR+})),
if we recall the replacement $\textrm{Re}E_{r}\rightarrow E_{s}$
and $E_{\textrm{YSR}}=\Delta$ as a result of the dissolution of the
YSR state into the upper branch single-particle states.

Finally, we present the spectrum across the BEC-BCS crossover as a
function of $1/(k_{F}a)$ in Fig. \ref{fig:NMPol}(c). Towards the
BEC side, we observe that the spectral weight of the singularity decreases
{[}which can be implied by the increase of $Z_{a}+Z_{r}$ shown in
Fig. \ref{fig:NMPol}(b){]}. Eventually, the singularity and the repulsive
polaron merges at around $1/(k_{F}a)\simeq0.5$, which coincides where
the chemical potential $\mu$ is changing from positive to negative.

\section{Experimental realization}

Our predictions could be readily examined in cold-atom experiments.
Indeed, several quantum mixtures consisting of a Fermi superfluid
and a Bose condensate have already been demonstrated, including $^{6}$Li-$^{7}$Li
\citep{Salomon2014Science}, $^{6}$Li-$^{41}$K \citep{YaoXingCan2016PRL},
and $^{6}$Li-$^{174}$Yb \citep{Roy2017PRL} mixtures. Quantum mixtures
such as $^{6}$Li-$^{133}$Cs \citep{Tung2013PRA}, $^{6}$Li-$^{168}$Er
\citep{Schafer2022PRA}, and $^{6}$Li-$^{168}$Er \citep{Schafer2022PRA}
should also be available soon, since the interspecies Feshbach resonances
have been characterized recently. In these mixtures, polaron physics
can be explored by reducing the concentration of the bosonic component.
For $^{6}$Li-$^{174}$Yb, $^{6}$Li-$^{133}$Cs, and $^{6}$Li-Er
systems, the minority bosonic species have different polarizability,
which allows imposing a deep optical lattice to localize the impurity
without affecting much the itinerant fermions. Even without the optical
lattice, our calculations still give quantitatively accurate predictions
due to the extremely large mass ratio. The response functions predicted
here can be measured via established methods: $S(t)$ can be accessed
via an interferometric Ramsey scheme; $A(\omega)$ can be obtained
in rf-spectroscopy.

As a concrete example, let us focus on the $^{6}$Li-$^{133}$Cs mixture.
Nowdays, a two-component Fermi superfluid of $^{6}$Li atoms in the
lowest two energy hyperfine states $\left|1,2\right\rangle =\left|F=1/2,m_{F}=\pm1/2\right\rangle $
is a typical setup to realize the BEC-BCS crossover in cold-atom laboratories,
owing to a broad Feshbach resonance at $B_{0}\simeq832$ G. The Feshbach
resonances between $^{133}$Cs and $^{6}$Li have been accurately
calibrated in 2013 \citep{Tung2013PRA}. Remarkably, in its lowest
energy state $\left|a\right\rangle =\left|F=3,m_{F}=3\right\rangle $
$^{133}$Cs atoms have a broad Feshbach resonance near $B_{0}$ with
$^{6}$Li atoms in both hyperfine states $\left|1,2\right\rangle $.
The resonances locate at $B_{0\uparrow}=843.4(2)$ G for $\left|\textrm{Li}:1\right\rangle +\left|\textrm{Cs}:a\right\rangle $
and $B_{0\downarrow}=889.0(2)$ G for $\left|\textrm{Li}:2\right\rangle +\left|\textrm{Cs}:a\right\rangle $.
The three closely located broad Feshbach resonances mean that we can
conveniently tune the magnetic field, to reach three significant scattering
lengths $a$, $a_{\uparrow}$ and $a_{\downarrow}$ at the same time.
In particular, by sweeping the magnetic field near $B_{0\uparrow}=843.4(2)$
G, the parameter sets used in Fig. \ref{fig:AwM} and Fig. \ref{fig:AwNM}
can be easily realized.

\section{Discussions and applications}

The present work shows how to generalize FDA to the system of a heavy
impurity immersed in a BCS superfluid. This formalism allows us to
construct an exact model to investigate polaron physics, which gives
all the universal polaron features, such as attractive and repulsive
polaron, dark continuum, and molecule-hole continuum. In our model,
the existence of polarons is protected from OC since the superfluid
pairing gap suppresses multiple particle-hole excitations, which plays
a similar role as the recoil energy of a mobile impurity in conventional
Fermi polarons. In addition, we have shown in an accompanying paper
\citep{AccompanyingShort2022PRL} that the pairing gap can also protect
the polarons from thermal fluctuations, allowing experimental studies
at a more accessible temperature $k_{B}T\sim\Delta$. Our results
for the non-magnetic impurity case also show some surprising results:
the existence of a repulsive polaron with an infinite lifetime and
an additional singularity. These peculiar characteristics only occur
at the perfect balance of the two scattering lengths, where the impurity
can only excite gapless density fluctuations. It would be interesting
to find an intuitive understanding of the underlying physics in future
studies.

Our predictions can also be applied to measure various exciting features
of the Fermi superfluid, although the BCS description is only quantitatively
reliable on the BCS side, and become only qualitatively reliable near
the unitary limit and the BEC side. In the magnetic impurity case,
the polaron spectrum at a finite but low temperature shows sharp features
that measure the sub-gap YSR bound states. In particular, if $E_{a}$,
${\rm Re}(E_{r})$, $E_{{\rm YSR}}^{(-)}$ and $E_{{\rm YSR}}^{(+)}$
shown in Fig. (\ref{fig:AwM})(d) are all measured accurately, Eqs.
(\ref{eq:EYSR-}) and (\ref{eq:EYSR+}) give rise to,
\begin{equation}
2\Delta=E_{a}+{\rm Re}(E_{r})-E_{{\rm YSR}}^{(-)}-E_{{\rm YSR}}^{(+)},
\end{equation}
independent on $E_{\textrm{YSR}}$. We believe that this relation
may only depend on the existence of a pairing gap and an in-gap bound
state, and therefore holds independent of the theoretical model used
in this work. This allows a highly accurate measurement of the pairing
gap $\Delta$ at the whole BEC-BCS crossover. In the non-magnetic
impurity case, there is also a finite temperature feature associated
with the singularity $E_{s}-2\Delta$, which can be applied to measure
the pairing gap $\Delta$.
\begin{acknowledgments}
We are grateful to Xing-Can Yao for insightful discussions. This research
was supported by the Australian Research Council's (ARC) Discovery
Program, Grants No. DE180100592 and No. DP190100815 (J.W.), and Grant
No. DP180102018 (X.-J.L). 
\end{acknowledgments}

\appendix

\section{The BCS-Leggett theory of the BEC-BCS crossover\label{sec:AppA_BCS}}

For a given scattering length $a$ and temperature $T$, $\Delta$
and $\mu$ are determined by the mean-field number and gap equations,
\begin{equation}
\sum_{\mathbf{k}}\left[1-\frac{\xi_{\mathbf{k}}}{E_{\mathbf{k}}}+2\frac{\xi_{\mathbf{k}}}{E_{\mathbf{k}}}f\left(E_{\mathbf{k}}\right)\right]=n,
\end{equation}
 
\begin{equation}
\frac{m}{4\pi\hbar^{2}a}+\sum_{\mathbf{k}}\left[\frac{1-2f\left(E_{\mathbf{k}}\right)}{2E_{\mathbf{k}}}-\frac{1}{2\epsilon_{\mathbf{k}}}\right]=0,
\end{equation}
where $f(E_{\mathbf{k}})=[\exp(-E_{\mathbf{k}}/k_{B}T)+1]^{-1}$ is
the Fermi--Dirac distribution, with $k_{B}$ is the Boltzmann constant.
Here $E_{\mathbf{k}}=\sqrt{\xi_{\mathbf{k}}^{2}+\Delta^{2}}$ are
the eigenvalues of Eq. (\ref{eq:hBCS}) with the corresponding eigenvector
$[u_{\mathbf{k}},v_{\mathbf{k}}]^{T}$, where $u_{\mathbf{k}}^{2}=\left[1+\xi_{\mathbf{k}}/E_{\mathbf{k}}\right]/2$,
$v_{\mathbf{k}}^{2}=1-u_{\mathbf{k}}^{2}$ and $2u_{\mathbf{k}}v_{\mathbf{k}}=\Delta/E_{\mathbf{k}}$.

\bibliography{RefHeavyCrossoverPolaron_long}

\end{document}